\documentclass[%
 reprint,nofootinbib,
superscriptaddress,
 amsmath,amssymb,
aps,
prd,
floatfix
]{revtex4-2}

\usepackage[colorlinks,linkcolor=red,anchorcolor=green,citecolor=blue,CJKbookmarks=True]{hyperref}
\usepackage{graphicx}
\usepackage{dcolumn}
\usepackage{bm}
\usepackage{color}
\usepackage{booktabs}
\usepackage{orcidlink}

\begin{document}

\preprint{APS/123-QED}

\title{Accelerating Black Hole Image Generation via Latent Space Diffusion Models}

\author{Ao Liu}
\altaffiliation{These authors contributed equally to this work.}
\affiliation{College of information science and engineering, Hunan Normal University, Changsha 410081, China.}

\author{Xudong Zhang}
\altaffiliation{These authors contributed equally to this work.}
\affiliation{Department of Physics, Institute of Interdisciplinary Studies, Key Laboratory of Low Dimensional Quantum Structures and Quantum Control of Ministry of Education, Synergetic Innovation Center for Quantum Effects and Applications, Hunan Normal University, Changsha 410081, China.}

\author{Lin Ding}
\affiliation{Department of Physics, Institute of Interdisciplinary Studies, Key Laboratory of Low Dimensional Quantum Structures and Quantum Control of Ministry of Education, Synergetic Innovation Center for Quantum Effects and Applications, Hunan Normal University, Changsha 410081, China.}

\author{Cuihong Wen\orcidlink{0000-0003-2668-4503}}
 \email{cuihongwen@hunnu.edu.cn}
\affiliation{College of information science and engineering, Hunan Normal University, Changsha 410081, China.}

\author{Wentao Liu\orcidlink{0009-0008-9257-8155}}
\email{wentaoliu@hunnu.edu.cn}
\affiliation{Lanzhou Center for Theoretical Physics, Key Laboratory of Theoretical Physics of Gansu Province, 
Key Laboratory of Quantum Theory and Applications of MoE,
Gansu Provincial Research Center for Basic Disciplines of Quantum Physics, 
Lanzhou University, Lanzhou 730000, China}
\affiliation{Institute of Theoretical Physics $\&$ Research Center of Gravitation,
Lanzhou University, Lanzhou 730000, China}

\author{Jieci Wang\orcidlink{0000-0001-5072-3096}}
 \email{jieciwang@hunnu.edu.cn}
\affiliation{Department of Physics, Institute of Interdisciplinary Studies, Key Laboratory of Low Dimensional Quantum Structures and Quantum Control of Ministry of Education, Synergetic Innovation Center for Quantum Effects and Applications, Hunan Normal University, Changsha 410081, China.}


\begin{abstract}
Interpreting horizon-scale black hole images currently relies on computationally intensive General Relativistic Ray Tracing (GRRT) simulations, which pose a significant bottleneck for rapid parameter exploration and high-precision tests of strong-field gravity. We demonstrate that physically accurate black hole images, synthesized from magnetized accretion flows, inherently reside on a low-dimensional manifold-encoding the essential features of spacetime geometry, plasma distribution, and relativistic emission. Leveraging this structure, we introduce a physics-conditioned diffusion model that operates in a compact latent space to generate high-fidelity black hole imagery directly from physical parameters. The model accurately reproduces critical observational signatures from full GRRT simulations-such as shadow diameter, photon-ring structure, and relativistic brightness asymmetry-while achieving over a fourfold reduction in computational expense. Compared with the previous generation of denoising diffusion models, the proposed approach achieves significant improvements in image quality, reconstruction fidelity, and parameter estimation accuracy, while reducing the average inference time per black hole image from 5.25 seconds to 1.15 seconds. Our work establishes diffusion-based latent models as efficient and scalable substitutes for traditional radiative transfer solvers, offering a practical framework toward real-time modeling and inference for next-generation black hole imaging.

\end{abstract}

\maketitle


\section{Introduction}
The historic imaging of M87* and Sgr A* by the Event Horizon Telescope (EHT) has transformed black holes from theoritical abstractions into observable entities \citep{2019ApJ...875L...1E, 2022ApJ...930L..12E, 2024A&A...681A..79E}. 
At the heart of these observations lies the black hole shadow, a dark central region bounded by a bright emission ring, which serves as a crucial probe for strong-field gravity and a testbed for general relativity \citep{2018NatAs...2..585M,Uniyal:2025uvc,Crinquand:2022ogr,Gao:2023mjb,Sui:2023tje,Liu:2024soc,Liu:2024iec,Liu:2024lbi,Liu:2024lve,Aslam:2025hgl,Wang:2025qpv,Zeng:2025pch,Huang:2025xqd,Wang:2025dfn,Liu:2025lwj}. 
To strictly interpret these observations, theoretical templates must be constructed to decode the observational fingerprints of the spacetime geometry \cite{2021ApJ...912...35N,Cunha:2018acu,Perlick:2021aok,Bronzwaer:2021lzo,Konoplya:2021slg,Vagnozzi:2022moj,Bambi:2019tjh,Chen:2022scf,Ayzenberg:2023hfw,Hou:2024qqo,Qiao:2025tnx}. 
However, the geometry itself is invisible and often described by non-integrable metrics where analytical symmetries are broken, necessitating advanced ray-tracing techniques \cite{Cunha:2015yba,Cunha:2019dwb,Hu:2020usx,Bacchini:2021fig,Liu:2025wwq,Zhang:2025xnl,Huang:2024gtu,Zhu:2025ouf}. 
Moreover, the shadow is revealed only when illuminated by the complex dynamics of surrounding plasma, typically modeled as magnetized accretion flows (RIAF) \citep{2014ARA&A..52..529Y, 2022MNRAS.511.3795N,Zhu:2025jqh,Zhu:2024vxw,Zhu:2023kei,Zeng:2021dlj}.
Bridging these geometric and physical requirements relies on General Relativistic Ray Tracing (GRRT) codes, which act as the essential link between abstract gravitational theories and EHT data \cite{Abramowicz:2011xu,EventHorizonTelescope:2019pcy,Cupp:2025zwv,Chen:2024nua}. 
By integrating photon trajectories in curved spacetime with the solution to the radiative transfer equation, GRRT synthesizes these physical models into essential observational signatures \cite{1973blho.conf..215B, 2000ApJ...528L..13F,Gralla:2019xty}.

However, the prohibitive cost of GRRT limits the extensive parameter surveys required for rigorous data comparison \citep{2020ApJ...897..139B, 2024MNRAS.535.3181Y, 2024IAUGA..32P1132C}.
Deep generative models offer a solution, with denoising diffusion probabilistic models emerging as superior to Generative Adversarial Networks and variational autoencoders due to their stable training and detailed mode coverage \citep{2026PatRe.17212577Z, 2026PatRe.17112232L, 2026PatRe.17112081X, 2026Meas..25919599O, 2026PatRe.16911934C}. 
Recently,  the Branch-Corrected Denoising Diffusion Model (BCDDM) has been proposed to synthesize physics-conditioned images \citep{2025ApJS..279...10L}. Compared with GRRT and other deep learning methods, BCDDM introduces a novel framework that substantially reduces the computational cost associated with generating black hole images while simultaneously improving parameter prediction accuracy.  However, the BCDDM relies on diffusion processes within the raw pixel space. 
This high-dimensional operation remains computationally demanding, preventing the leap to truly real-time generation.

The manifold hypothesis, which posits that high-dimensional data such as black hole images reside on a low-dimensional latent manifold \citep{2013arXiv1310.0425F, 6789755, 2012arXiv1206.5538B, 2021arXiv210408894P}, provides a foundational principle for addressing the dimensionality bottleneck. In this context, Principal Component Analysis (PCA) has been effectively employed to extract a compact eigenimage basis from simulations to characterize image variability and constrain reconstructions \citep{2018ApJ...864....7M, 2022ApJ...927..111H, 2025ApJ...984...86P}. However, the inherent linearity of PCA restricts its expressive power, hindering its ability to accurately model the complex, nonlinear geometry of the data manifold and to generate novel, high-fidelity samples. To overcome this critical limitation and more faithfully capture the manifold's geometry, we propose the Latent Self-Attentive Denoising Diffusion Model (LSA-DDM), which is a novel generative framework designed for both computational efficiency and physical fidelity. Such model leverages a nonlinear diffusion process to learn the data distribution directly on the latent manifold, augmented by a self-attention mechanism to capture complex global dependencies. This approach establishes a more powerful and flexible generative framework that surpasses the limitations inherent to linear methods.

The proposed framework centers on a two-stage methodology: first, we leverage PCA to distill high-dimensional black hole images into a compact latent representation that preserves the most physically salient morphological variations. By then training a conditional diffusion model entirely within this low-dimensional space, we fundamentally overcome the computational bottleneck inherent to pixel-space generation. One more key innovation is the integration of a self-attention mechanism directly into the parameter-conditioning pathway, which empowers the model to capture the complex, nonlinear relationships among accretion parameters with high fidelity. As a result, the LSA-DDM not only achieves a substantial acceleration in image generation but does so while uncompromisingly preserving the physical accuracy and parameter sensitivity of the original GRRT training set. This establishes a new paradigm for efficient, physically-grounded generative modeling, paving the way for rapid simulation, robust data augmentation, and advanced parameter estimation in black hole imaging.

\section{Physical Image Generation Problem and Dataset Construction}

\subsection{Overall Framework}

\begin{figure*}[t] 
\centering 
\includegraphics[width=0.8\textwidth]{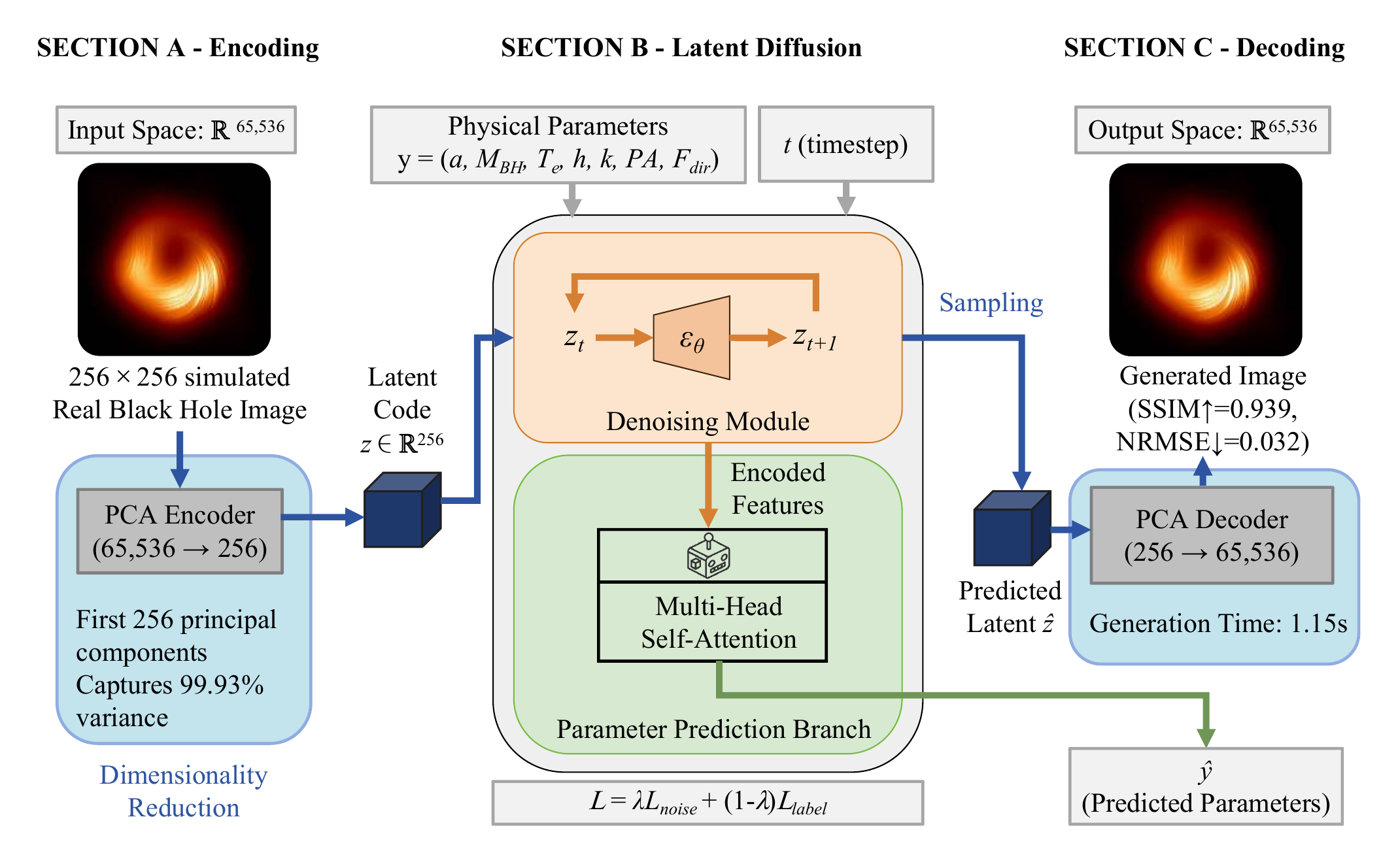} %
\caption{Overall framework of the proposed Latent Self-Attentive Denoising Diffusion Model. This figure includes the training process and image generation process.}
\label{fig:overall_framework}
\end{figure*}

Figure \ref{fig:overall_framework} illustrates the dual-pathway framework of the proposed LSA-DDM. The model operates in two distinct modes: a training process that learns the mapping from images to parameters within a compact latent space, and a generation (sampling) process that generates images from arbitrary physical parameters.

During training, the pipeline begins with a high-resolution GRRT-simulated black hole image (256$\times$256 pixels). This image is first compressed by a fixed PCA Encoder into a low-dimensional latent vector $\mathbf{z}$. Concurrently, the corresponding set of ground-truth physical parameters (labels) $\mathbf{y}$ is extracted, and for each training instance, a random timestep $t$ is sampled from a uniform distribution over ${1, ..., T}$. Both the latent vector $\mathbf{z}$ and the parameters $\mathbf{y}$ are fed into the core network. The model, featuring a parameter prediction branch augmented with a self-attention mechanism, is trained to perform two tasks simultaneously: predicting the noise component $\epsilon$ to denoise the latent code, and regressing the physical parameters $\mathbf{\hat{y}}$. The total loss is computed from the differences between the predicted and target noise, as well as between the predicted $\mathbf{\hat{y}}$ and true $\mathbf{y}$ parameters, enabling the model to learn a physically-consistent latent manifold.

For synthesis, the process is conditioned solely on a target set of physical parameters $\mathbf{y}$. Starting from a random Gaussian noise vector in the latent space, the trained LSA-DDM iteratively denoises it over multiple steps, guided by the conditioning parameters $\mathbf{y}$. The model outputs both the denoised latent vector $\mathbf{\hat{z}}$ and the corresponding predicted parameters $\mathbf{\hat{y}}$. Finally, the latent vector $\mathbf{\hat{z}}$ is projected back to the high-dimensional pixel space via the fixed PCA Decoder to yield the final generated black hole image $\mathbf{\hat{x}}$. This decoupled design allows efficient, high-fidelity image synthesis entirely within a low-dimensional, semantically structured latent space, driven by explicit physical constraints.

\subsection{Latent Space Construction with Principal Component Analysis}

The generation of high-resolution black hole images, each composed of $256 \times 256$ pixels, operates within a $65,\!536$-dimensional pixel space. Direct application of diffusion models in this high-dimensional space proves computationally intensive and suffers from data inefficiency. To overcome this curse of dimensionality, we employ PCA to construct a compact, informative latent representation that captures the essential morphological features of the images~\cite{2025PhRvD.111f3523P}.

The mathematical framework for PCA presented here follows the standard formulation found in authoritative machine learning literature~\cite{bishop2006pattern, goodfellow2016deep}. Given a preprocessed dataset of $n$ black hole images, each reshaped into a column vector $\mathbf{x}_i \in \mathbb{R}^{D}$, we form the data matrix $\mathbf{X} = [\mathbf{x}_1, \mathbf{x}_2, ..., \mathbf{x}_n] \in \mathbb{R}^{D \times n}$. The PCA seeks an orthogonal projection that maximizes the variance of the projected data. This is achieved by solving the eigenvalue decomposition of the sample covariance matrix $\mathbf{C}$
\begin{equation}
    \mathbf{C} = \frac{1}{n-1} \sum_{i=1}^{n} (\mathbf{x}_i - \bar{\mathbf{x}})(\mathbf{x}_i - \bar{\mathbf{x}})^T = \mathbf{V} \mathbf{\Lambda} \mathbf{V}^T,
    \label{eq:covariance_decomp}
\end{equation}
where $\bar{\mathbf{x}} = \frac{1}{n} \sum_{i=1}^{n} \mathbf{x}_i$ is the mean image, $\mathbf{V} \in \mathbb{R}^{D \times D}$ is an orthogonal matrix whose columns $\{\mathbf{v}_1, \mathbf{v}_2, ..., \mathbf{v}_D\}$ are the principal components (eigenvectors), and $\mathbf{\Lambda} = \operatorname{diag}(\lambda_1, \lambda_2, ..., \lambda_D)$ is a diagonal matrix of eigenvalues ordered such that $\lambda_1 \geq \lambda_2 \geq ... \geq \lambda_D \geq 0$. 
The eigenvalue $\lambda_j$ corresponds to the variance captured along its associated direction $\mathbf{v}_j$.

The intrinsic dimensionality governing the structure of $256 \times 256$ black hole image data is significantly lower than the 65,536 dimensions of the original pixel space, a fact clearly demonstrated by the rapid decay of eigenvalues discussed in detail in Section \ref{pca_analysis}.
Our analysis of the cumulative explained variance shows that retaining the first $d = 256$ principal components achieves an optimal balance between high compression and minimal information loss. This defines our latent space $\mathcal{Z} = \mathbb{R}^{d}$. The encoding of an image $\mathbf{x}$ into a latent code $\mathbf{z} \in \mathcal{Z}$ is a linear projection
\begin{equation}
\begin{aligned}
  \mathbf{z} = &\mathbf{W}_{\text{enc}} (\mathbf{x} - \bar{\mathbf{x}}), \\
 &\mathbf{W}_{\text{enc}} = [\mathbf{v}_1, \mathbf{v}_2, ..., \mathbf{v}_d]^T \in \mathbb{R}^{d \times D}.
    \label{eq:encoding}
\end{aligned}
\end{equation}
Conversely, the decoding (reconstruction) from the latent space back to an approximation in the pixel space is given by
\begin{equation}
\begin{aligned}
\hat{\mathbf{x}} = &\mathbf{W}_{\text{dec}} \mathbf{z} + \bar{\mathbf{x}}, \\
&\mathbf{W}_{\text{dec}} = [\mathbf{v}_1, \mathbf{v}_2, ..., \mathbf{v}_d] \in \mathbb{R}^{D \times d},
    \label{eq:decoding}
\end{aligned}
\end{equation}
Here, $\mathbf{W}_{\text{dec}} = \mathbf{W}_{\text{enc}}^T$, and the mean image $\bar{\mathbf{x}}$ is added back. 
The pair $(\mathbf{W}_{\text{enc}}, \mathbf{W}_{\text{dec}}, \bar{\mathbf{x}})$ forms a fixed, non-trainable autoencoder. 
The mean squared reconstruction error for this truncation is minimized and equals the sum of the discarded eigenvalues: $\sum_{j=d+1}^{D} \lambda_j$.

The core of our generative framework is a conditional denoising diffusion model that operates within the PCA-derived latent space $\mathcal{Z}$. The model learns to reverse a forward diffusion process, gradually denoising a latent code $\mathbf{z}_T \sim \mathcal{N}(0, \mathbf{I})$ over $T$ steps to produce a clean latent representation $\mathbf{z}_0$ (denoted as $\hat{\mathbf{z}}$), which corresponds to a reconstructed black hole image conditioned on a set of physical parameters $\mathbf{y}$.

Our latent diffusion process operates within a 256-dimensional vector space, rather than on the original two-dimensional images. To accommodate this one-dimensional domain, we re-engineered the standard U-Net architecture. Specifically, all spatial operations, including convolutions, pooling, and transposed convolutions, were replaced with their one-dimensional counterparts. While the dimensionality of the operations has changed, the fundamental U-Net design is preserved. This includes the symmetric encoder-decoder structure and the crucial skip connections that link them. This architecture treats the latent vector as a 256-step sequence, allowing the model to effectively capture multi-scale features and hierarchical dependencies along the latent manifold. Ultimately, this design enables the efficient modeling of complex structural information corresponding to features in the original high-dimensional image space. 

A central innovation of our model is the integration of a self-attention mechanism into the parameter prediction branch, designed to actively enhance the correspondence between the extracted image features and the target physical parameters $\mathbf{y}$. This branch processes the high-dimensional feature maps from the encoder to predict the associated physical conditions. While standard feed-forward layers can establish a baseline mapping, we introduce self-attention to explicitly model and strengthen the long-range dependencies and complex interactions among different feature dimensions.

\begin{figure}[htbp] 
\centering 
\includegraphics[width=0.5\textwidth]{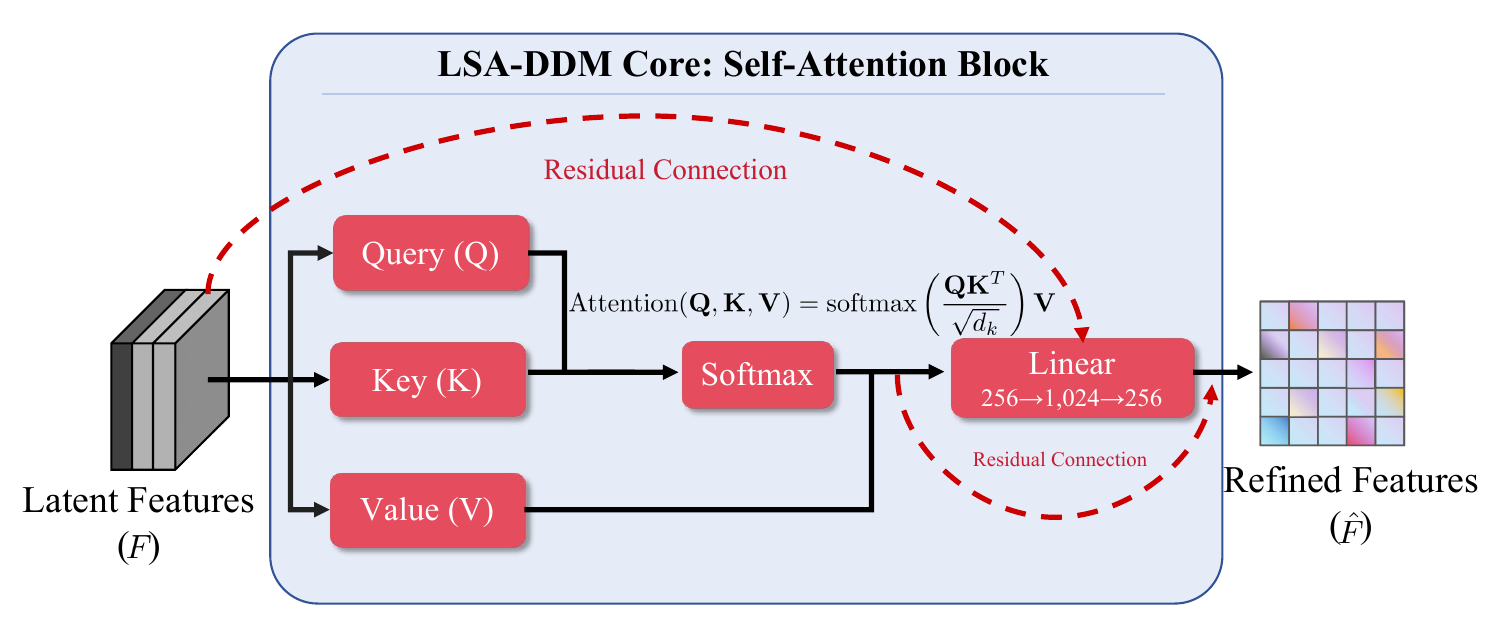} %
\caption{Schematic of the self-attention block integrated into the parameter prediction branch. The diagram illustrates the flow from input features through linear projections to Query (Q), Key (K), and Value (V) vectors, the computation of attention weights, and the final weighted aggregation that produces a refined feature representation. This module enables the model to capture long-range dependencies within the feature set, enhancing physical consistency.}
\label{fig4} 
\end{figure}

A Self-Attention block is incorporated after an initial feature projection within the parameter prediction branch, as shown in Fig. \ref{fig4}. 
This mechanism allows the network to dynamically re-weight and contextualize features by computing interactions across all spatial or feature-channel locations. For a given feature representation $\mathbf{F}$, the self-attention operation is defined as~\cite{vaswani2017attention}
\begin{equation}
    \text{Attention}(\mathbf{Q}, \mathbf{K}, \mathbf{V}) = \text{softmax}\left(\frac{\mathbf{Q}\mathbf{K}^T}{\sqrt{d_k}}\right)\mathbf{V},
\end{equation}
where $\mathbf{Q}, \mathbf{K}, \mathbf{V}$ are linear projections of $\mathbf{F}$, and $d_k$ is the dimensionality of the key vectors. Employing multi-head attention enables the model to jointly attend to information from different representation subspaces. This design fosters a more cohesive and internally consistent feature integration, thereby sharpening the model's ability to condition the generated latent codes on a precise and physically coherent parameter set $\mathbf{y}$. The enhanced conditioning signal guides the subsequent denoising U-Net pathway more effectively, leading to generated images with higher fidelity to the specified physical constraints.

The integration of self-attention into the parameter prediction branch is motivated by the need for tightened conditioning. By allowing the network to dynamically re-weight and contextualize features based on their global interactions, the module learns a more robust and internally consistent mapping from image features to parameters. This results in two principal benefits for the overall diffusion model: (1) It provides a more accurate and physically grounded conditional signal $\mathbf{y}$ to guide the denoising U-Net pathway at every step, and (2) It drives the performance of the parameter prediction by directly supervising it through an auxiliary loss term. Consequently, the latent codes generated by the model are not only statistically plausible but are also strongly anchored to the input physics, ensuring that the final decoded images exhibit high physical fidelity and consistency with the specified black hole properties.

\subsection{Black Hole Image Dataset and Preprocessing}
To train and evaluate the proposed latent space diffusion model, we employ a dataset of simulated black hole images that are consistent with the observational characteristics of M87* as captured by the EHT. These images are generated using the well-established GRRT method within the RIAF model framework, which is suitable for modeling the hot, optically thin plasma around low-accretion-rate black holes such as M87* and Sgr A*. The image dataset used in this work was publicly released by~\citet{Zhang111}. Details of the data generation process are provided in Appendix~\ref{Appendix_A}.

The GRRT simulations were performed using the \texttt{ipole} code to calculate synchrotron radiation at an observational frequency of 230 GHz. Each image in the dataset has a resolution of $256 \times 256$ pixels, covering a field of view of $160 \times 160$ micro-arcseconds, matching the scale of EHT observations. The black hole spin, mass, and key accretion flow properties are varied across simulations to create a diverse dataset. Specifically, seven physical parameters---including black hole spin ($a$), mass ($M_{\text{BH}}$), electron temperature ($T_e$), accretion disk scale height ($h_{\text{disk}}$), Keplerian factor ($k$), position angle (PA), and fluid rotation direction ($F_{\text{dir}}$)---are sampled within their physically plausible ranges (see, e.g., Table 1 of \citep{2025ApJS..279...10L} for detailed bounds). The observer inclination is fixed at $163^\circ$, and the total flux density is normalized to 0.5 Jy for all images, consistent with the M87* case. The final curated dataset comprises 2157 distinct black hole images, each uniquely mapped to its corresponding set of seven physical parameters.

Effective preprocessing is crucial for stabilizing the training of deep generative models and for the subsequent PCA. We apply a two-stage normalization procedure to both the physical parameters and the image pixels.

The physical parameters span several orders of magnitude (e.g., mass and temperature). To prevent numerical instabilities and ensure all features contribute equally to the conditioning input of our model, we apply Z-score normalization to each parameter. For a parameter $p$ with mean $\mu_p$ and standard deviation $\sigma_p$ over the dataset, the normalized value $\tilde{p}$ is given by
\begin{equation}
\tilde{p} = \frac{p - \mu_p}{\sigma_p},
\end{equation}
this linear transformation results in a parameter distribution with zero mean and unit variance. Prior to normalization, the electron temperature $T_e$ is log-transformed due to its log-normal characteristic span across orders of magnitude.

Each simulated image, represented as a flux density matrix $\mathbf{S}$, is normalized to the range [0, 1] by dividing each pixel value by the maximum flux density $S_{\text{max}}$ of that specific image:
\begin{equation}
\mathbf{S}_{\text{norm}} = \frac{\mathbf{S}}{S_{\text{max}}}, \quad \text{where } S_{\text{max}} = \max(\mathbf{S}).
\end{equation}
This pixel-wise scaling preserves the relative structure and morphology of the black hole shadow, photon ring, and emission region while providing a consistent input scale for the PCA transformation. Since all original GRRT images are calibrated to the same total flux (0.5 Jy), this normalization is reversible, allowing generated images to be rescaled back to physically meaningful flux units.

\section{Experiments and Results}
\subsection{Experimental Setup}
All experiments were performed on a single NVIDIA RTX 3090 GPU. The proposed diffusion model adopts a linear noise schedule, defined by $\beta_{\text{start}}=1\times10^{-4}$ to $\beta_{\text{end}}=2\times10^{-2}$ over $T=1,000$ diffusion steps. A core architectural contribution is the incorporation of an 8‑head self‑attention mechanism into the parameter‑prediction branch, designed to capture global feature dependencies and thereby enhance the physical consistency of generated images. The dataset was randomly split into training, validation, and test subsets in an 8:1:1 ratio. The model was trained for 10000 epochs with a batch size of 32, optimized using Adam (initial learning rate $0.005$, weight decay $10^{-4}$) together with a cosine annealing scheduler. The total loss function combines the diffusion denoising loss and a parameter‑prediction loss, with a balancing weight set to $\lambda=0.5$. Model performance was evaluated on the held‑out test set based on image quality measured by the normalized root mean square error (NRMSE) and structural similarity index (SSIM), the parameter‑prediction accuracy measured by the mean absolute error (MAE), and single‑image generation speed. All experiments were conducted with a fixed random seed 42 to ensure reproducibility.

\subsection{Analysis of the PCA Latent Space}\label{pca_analysis}

\begin{figure*}[htbp] 
\centering 
\includegraphics[width=1\textwidth]{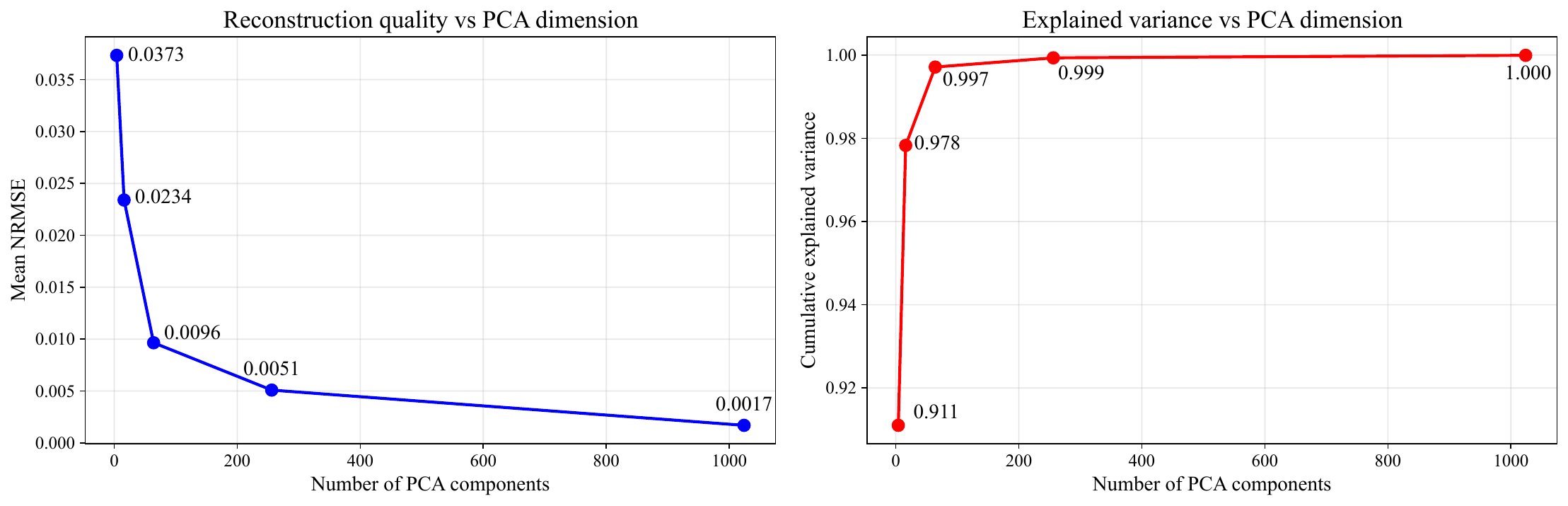} %
\caption{Reconstruction quality assessment across varying PCA dimensions. As the dimensionality increases, the NRMSE decreases and visual fidelity improves sharply until a saturation point near $d=256$. The minimal improvement beyond 256 components confirms that $d=256$ represents the optimal trade-off between fidelity and efficiency.}
\label{fig:recon_quality} 
\end{figure*}

The determination of the latent space dimension $d$ balances empirical reconstruction quality with theoretical variance retention and generation efficiency. As shown in Fig. \ref{fig:recon_quality}, reconstructions using a low number of principal components (e.g., \(d=4, 16, 64\)) exhibit high mean NRMSE and a significant loss of fine structural details. The reconstruction quality improves markedly as \(d\) increases to 256, and reaches a plateau near \(d=256\). Quantitatively, the improvement in NRMSE from \(d=256\) to \(d=1024\) is merely 0.0034, demonstrating rapidly diminishing returns. The reconstructions at \(d=256\) and \(d=1024\) are visually nearly indistinguishable and both correspond to a very high cumulative explained variance (\(>0.999\)). This suggests that $d=256$ is a pragmatic saturation point, optimally capturing the critical physical information without introducing computational redundancy.

Formally, the proportion of total variance retained by the top-$d$ components is given by
\begin{equation}
R(d) = \frac{\sum_{j=1}^{d} \lambda_j}{\sum_{j=1}^{D} \lambda_j}.
\end{equation}

\begin{figure}[htbp] 
\centering 
\includegraphics[width=0.5\textwidth]{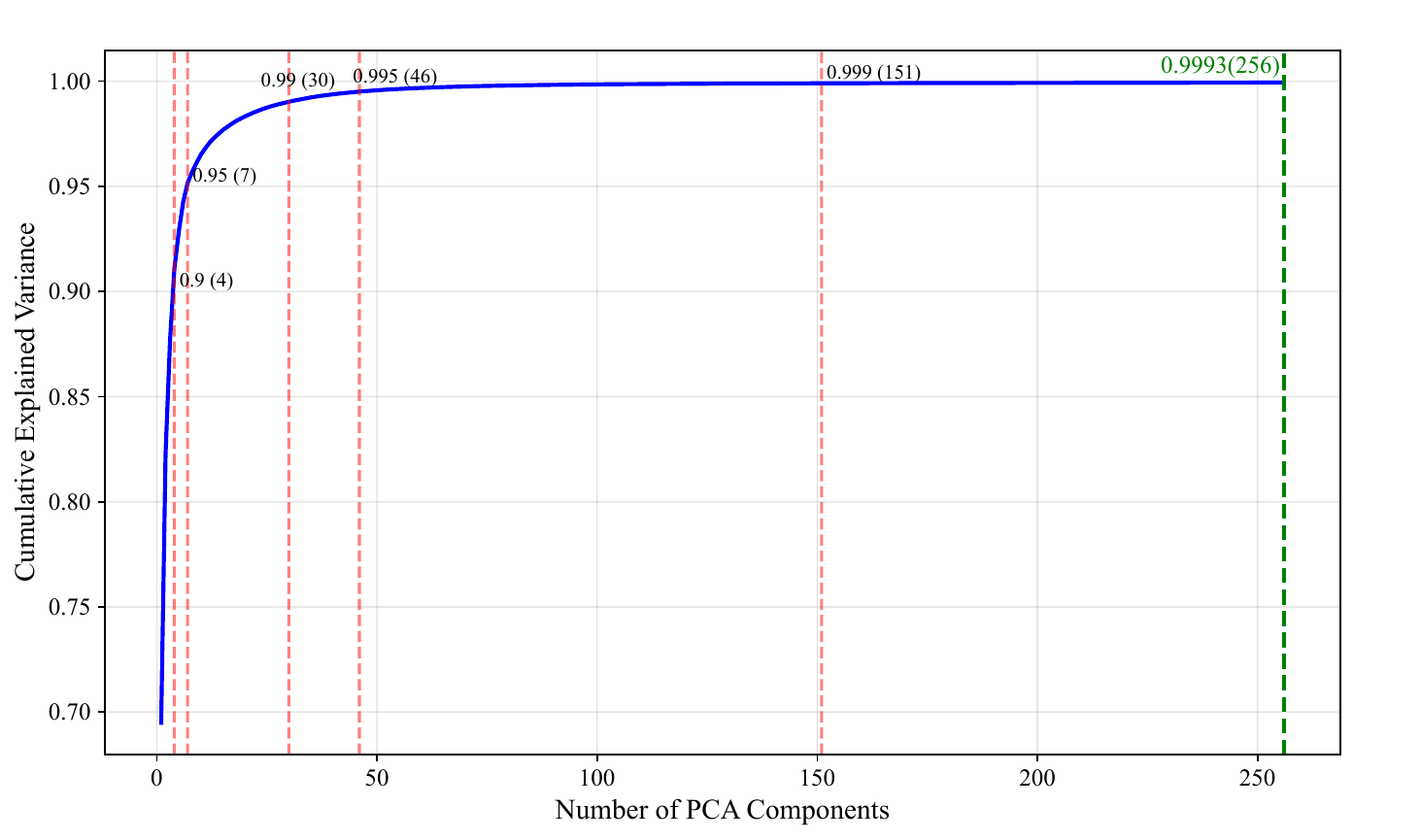} %
\caption{Cumulative explained variance as a function of the number of retained principal components. The first 151 components capture $99.9\%$ of the total variance. We select $d=256$ (retaining $R\approx0.9993$) to ensure compatibility with the U-Net's power-of-two architecture constraints, achieving a 256-fold compression from the original $65,536$ pixel dimensions.}
\label{fig:cumu_var} 
\end{figure}

As shown in Fig. \ref{fig:cumu_var}, analysis of the cumulative explained variance reveals that the first $d' = 151$ principal components are sufficient to capture $99.9\%$ ($R(d') = 0.999$) of the total variance in the dataset, indicating a very low intrinsic dimensionality. 
However, to seamlessly integrate with the subsequent U-Net-based diffusion model—which utilizes repetitive downsampling and upsampling operations that are most efficient when feature map dimensions are powers of two—we select the nearest power-of-two dimension. 
Therefore, we set $d = 256$. 
This choice yields a marginally higher variance retention of $R(256) \approx 0.9993$, while ensuring computational efficiency and compatibility within the deep learning framework. 
The resultant compression from the original $D = 65,\!536$ pixel dimensions to $d = 256$ (a factor of 256) facilitates training stability and accelerates generation by allowing the diffusion process to operate within a smooth, low-dimensional manifold that encodes the essential physical structures.

\subsection{Performance Benchmarking}
To quantitatively evaluate the efficiency gains of our proposed PCA-based latent diffusion framework, we compare its key performance metrics against the baseline BCDDM model which operates directly in the pixel space. The comparison is conducted under identical hardware and software environments, using the same RIAF black hole image dataset. The results, summarized in Table~\ref{tab:performance_benchmark}, highlight the substantial improvements achieved by our method.

\begin{table}[htbp]
    \centering
    \caption{Performance comparison between the baseline BCDDM and the proposed LSA-DDM.}
    \label{tab:performance_benchmark}
    \begin{tabular}{lcc}
        \toprule
        \textbf{Metric} & \textbf{BCDDM} & \textbf{LSA-DDM} \\
        \midrule
        Effective Latent Dimension & 65,536 & 256 \\
        Model Size (Parameters) & 247.09 M & $\mathbf{60.59}$ M \\
        Training Time per Epoch & 47.76 s & 13.37 s \\
        Image Generation Time & 5.25 s & $\mathbf{1.15}$ s \\
        \bottomrule
    \end{tabular}
\end{table}

The dramatic reduction in model size is primarily attributed to the U-Net architecture being redesigned to process 256-dimensional latent vectors, rather than full-resolution $256\times256$ images. This directly translates to faster training and lower memory footprint. The average time to generate a single black hole image is reduced from 5.25 seconds to approximately 1.15 seconds. This efficiency gain is the direct consequence of shifting the computationally intensive denoising diffusion process from a 65,536-dimensional pixel space to a compact 256-dimensional latent manifold.

\subsection{Ablation Study on PCA and Self-Attention}
\label{subsec:ablation}

To evaluate the individual contributions of PCA dimensionality reduction and the self-attention mechanism, we conducted an ablation study comparing three variants of our model: baseline model, BCDDM enhanced with PCA integration, and LSA-DDM. All models were assessed on image quality metrics (NRMSE, SSIM) and parameter prediction accuracy (MAE).

\begin{table*}[htbp]
    \centering
    \caption{Ablation study results. PCA integration improves image quality (↓NRMSE, ↑SSIM), and the addition of self-attention further enhances both reconstruction fidelity and parameter estimation accuracy.}
    \label{tab:ablation_study}
    \begin{tabular}{lccc}
        \toprule
        \textbf{Variant} & \textbf{NRMSE} $\downarrow$ & \textbf{SSIM} $\uparrow$ & \textbf{MAE} $\downarrow$ \\
        \midrule
        BCDDM       & 0.043\; ± \;0.022 & 0.925 \;±\; 0.039  & 0.082\;±\;0.039 \\
        BCDDM + PCA              & 0.059\;±\;0.022 & 0.881\;±\;0.047  & 0.171 \;±\; 0.099 \\
        \textbf{LSA-DDM} & \textbf{0.032\;±\;0.015} & \textbf{0.939\;±\;0.027} & \textbf{0.059\;±\;0.021} \\
        \bottomrule
    \end{tabular}
\end{table*}

\begin{figure*}[t] 
\centering 
\includegraphics[width=1\textwidth]{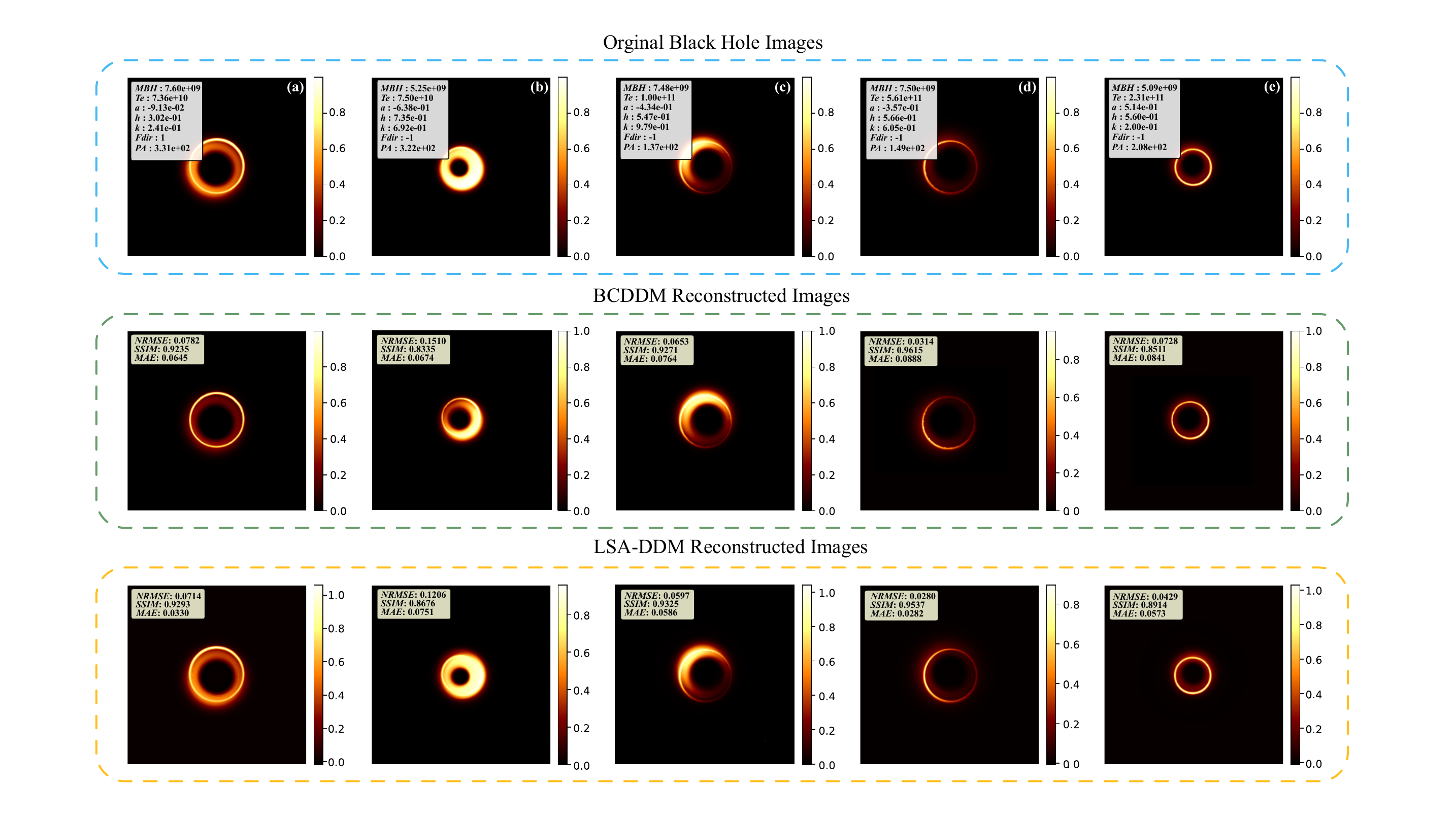} %
\caption{Visual comparison of black hole images generated by different model variants under identical physical parameters. Each panel displays, from top to bottom: the normalized ground-truth GRRT image, the image generated by the baseline BCDDM, the image from the PCA-enhanced BCDDM, and the image synthesized by the full proposed LSA-DDM. This side-by-side comparison qualitatively demonstrates the progressive improvement in visual fidelity achieved by our synergistic design.}
\label{fig:generated_samples} 
\end{figure*}
As shown in Table~\ref{tab:ablation_study}, the integration of PCA with the baseline BCDDM introduces a clear trade-off between efficiency and fidelity. While PCA enables the dramatic acceleration, it incurs an inevitable cost in reconstruction precision. Transitioning from BCDDM to the PCA-enhanced variant leads to a degradation in image quality, with NRMSE increasing from $0.043 $ to $0.059$ and SSIM decreasing from $0.925$ to $0.881$. Meanwhile, the compression inherent to PCA disrupts the feature manifold used for parameter regression, causing parameter prediction accuracy to substantially worsen, as evidenced by the MAE increasing from $0.082$ to $0.171$.

The proposed LSA-DDM, which incorporates a self-attention mechanism atop the PCA framework, successfully mitigates these drawbacks and achieves a significant improvement in performance. The self-attention module learns to dynamically re-weight and integrate the compressed latent features, effectively recovering semantically meaningful associations that were attenuated by PCA. This results in a more stable and robust conditioning process, where the input physical parameters can guide the denoising pathway more precisely. Consequently, LSA-DDM not only recovers the losses induced by PCA but surpasses the original BCDDM, achieving the best scores across all metrics: the lowest NRMSE ($0.032$), the highest SSIM ($0.939$), and the most accurate parameter prediction (MAE of $0.059$).

This ablation study validates a synergistic design where PCA enables efficiency and self-attention ensures fidelity. Visual evidence of this progressive improvement is provided in Fig. \ref{fig:generated_samples}, which showcases a comparative set of black hole images generated under identical physical parameters by the three model variants: the baseline BCDDM, the PCA-enhanced BCDDM, and the full proposed LSA-DDM.

\section{Conclusion}

In this work, we have introduced the LSA-DDM, a novel framework designed to accelerate the generation of black hole images by operating within a low-dimensional latent space. 
By leveraging PCA, we compress high-resolution black hole images into a compact 256-dimensional manifold that retains the essential physical and morphological features while discarding redundant pixel-level noise. 
This dimensionality reduction shifts the computationally intensive diffusion process from a $65,536$-dimensional pixel space to a smooth, low-dimensional latent space, resulting in a dramatic reduction in model size, training time, and generation time.

Our proposed model integrates a self-attention mechanism into the parameter-conditioning branch, which enhances the model's ability to capture complex, long-range dependencies among physical parameters, thereby improving the physical consistency and fidelity of the generated images. Experimental results show that the LSA-DDM achieves a generation speed of approximately 1.15 seconds per image, more than four times faster than the pixel-space diffusion model BCDDM, while maintaining high image quality, as measured by NRMSE and SSIM. Moreover, the model exhibits improved parameter prediction accuracy, demonstrating its ability to learn a robust mapping between image features and physical parameters.

The efficiency gains offered by LSA-DDM make it a practical tool for rapid data augmentation, parameter estimation, and model fitting in black hole astrophysics. While the current model is trained on images generated under the RIAF framework, the methodology is general and can be extended to other accretion flow models, including those incorporating jet emission or polarization information. Future work may explore nonlinear dimensionality reduction techniques, such as variational autoencoders, to further optimize the latent space representation, and incorporate multi-channel inputs to handle polarized black hole images, thereby providing even more comprehensive support for the analysis of EHT observations.

\begin{acknowledgments}

This work was supported by the National Natural Science Foundation of China (Grants No. 12374408, No. 12475051, and No. 12547147), the China Postdoctoral Science Foundation (Grant No. 2025M783393), and the Hunan province college students research learning and innovative experiment project (Grant No. 202510542241).
~\\~\\~\\~

\end{acknowledgments}

\appendix
\section{Physical Background of Black Hole Imaging}\label{Appendix_A}
We briefly review the canonical case in general relativity, namely the Kerr black hole spacetime, whose metric in Boyer-Lindquist coordinates is given by
\begin{equation}
\begin{aligned}
ds^2=&-dt^2+\Sigma d\theta^2 +\left( r^2+a^2 \right)\sin^2\theta d\varphi^2\\
&+ \frac{\Sigma}{\Delta}dr+\frac{2Mr}{\Sigma}\left( dt^2-a\sin^2\theta d\varphi \right)^2,
\end{aligned}
\end{equation}
where
\begin{equation}
\Delta=r^2-2Mr+a^2,\quad\quad \Sigma=r^2+a^2\cos^2\theta.
\end{equation}
Note that we have set $c=G=1$ and $M$ is the mass of the black hole.
To obtain images of a Kerr black hole surrounded by a geometrically thick, magnetized equilibrium torus, we consider general relativistic magnetohydrodynamics in curved spacetime, which is governed by the following three conservation equations~\cite{Komissarov:2006nz}
\begin{equation}
\nabla_\nu(\rho u^\mu)=0,\quad  \nabla_\mu T^{\mu\nu}=0,\quad   \nabla_\nu \bar{F}^{\mu\nu}=0.
\end{equation}
Here $u^\mu$ denotes the fluid four-velocity, $\rho$ is the rest-mass density, $F^{\mu\nu}$ is the electromagnetic field tensor, and $\bar{F}^{\mu\nu}$ is its dual.
The total energy-momentum tensor $T^{\mu\nu}$ receives contributions from both the fluid and the electromagnetic field, $T^{\mu\nu} = T^{\mu\nu}_{f} + T^{\mu\nu}_{\mathrm{EM}}$.
In general relativistic magnetohydrodynamics, the total energy-momentum tensor, where the enthalpy density of the fluid is denoted by $h$, can be written as
\begin{equation}
T^{\mu\nu} = (h + b^2) u^\mu u^\nu + \left(p + \frac{b^2}{2}\right) g^{\mu\nu} - b^\mu b^\nu ,
\end{equation}
where $p$ is the gas pressure and $b^\mu$ the magnetic four-vector in the comoving frame; the magnetic pressure is
$p_{\rm m}\equiv b^2/2$.

Following \cite{Komissarov:2006nz}, we impose stationarity and axisymmetry, $\partial_t=\partial_\varphi=0$, and assume purely circular motion together with a purely toroidal magnetic field, $u^\mu=(u^t,0,0,u^\varphi)$ and $b^\mu=(b^t,0,0,b^\varphi)$.
We further define the angular velocity $\Omega\equiv u^\varphi/u^t$ and the specific angular momentum $\ell\equiv -u_\varphi/u_t$, and adopt the constant angular-momentum prescription $\ell=\ell_0$. 
Under these assumptions, the Euler equation obtained from $\nabla_\mu T^{\mu\nu}=0$ reduces to an integrable form.
With the above specialization, the spatial components of the momentum equation can be written as \cite{Komissarov:2006nz}
\begin{equation}
\label{eq:euler_diff_tilde}
\partial_\nu \ln |u_t| -\frac{\Omega}{1-\ell\Omega}\,\partial_\nu \ell +\frac{\partial_\nu p}{h} +\frac{\partial_\nu \tilde p_{\rm m}}{\tilde h}=0,
\end{equation}
where the ``tilded'' quantities absorb the standard geometric factor
\begin{equation}
\label{eq:tildes}
\tilde h\equiv \mathcal{L}\,h,\qquad
\tilde p_{\rm m}\equiv \mathcal{L}\,p_{\rm m},\qquad
\mathcal{L}\equiv g_{t\varphi}^2-g_{tt}g_{\varphi\varphi}.
\end{equation}
For constant $\ell=\ell_0$, Eq.~\eqref{eq:euler_diff_tilde} can be integrated to
\begin{equation}
\label{eq:integrated_euler}
W-W_{\rm in}+\frac{\kappa}{\kappa-1}\frac{p}{h}+\frac{\eta}{\eta-1}\frac{p_{\rm m}}{h}=0,
\end{equation}
where $W_{\rm in}$ denotes the potential at the inner edge of the torus. 
The effective potential is defined by
\begin{equation}
\label{eq:W_def}
W(r,\theta)=\frac{1}{2}\ln\left|\frac{\mathcal{L}}{\mathcal{A}}\right|,
\qquad
\mathcal{A}\equiv g_{\varphi\varphi}+2\ell_0 g_{t\varphi}+\ell_0^2 g_{tt}.
\end{equation}

To close the system, we assume polytropic relations for the gas and magnetic pressures
\begin{equation}
\label{eq:polytropes}
p=K h^\kappa,\qquad p_{\rm m}=K_{\rm m} h^\eta,
\end{equation}
with constants $K,K_{\rm m},\kappa,\eta$. 
The relative magnetic strength is quantified by the plasma magnetization parameter $\beta\equiv p/p_{\rm m}$; in practice, fixing its value at the torus center sets the overall normalization of the magnetic field once the thermodynamic structure is specified. 
The stationary equilibrium solution  provides a self-consistent plasma configuration in Kerr spacetime, which serves as the physical background for subsequent image synthesis.
Given an electron-thermodynamics prescription \cite{Yuan:2014gma}, local emissivities and absorptivities can be assigned and general relativistic radiative transfer along null geodesics can be performed to synthesize images. 
We note that adopting more refined prescriptions for the electron temperature, such as the radially dependent power-law profile proposed in the Hou Disk model \cite{Hou:2023bep,Zhang:2024lsf,Wan:2025gbm}, can further improve the internal consistency of the simulated images, although the present framework remains fully applicable. 

\begin{figure}[t]
\centering
\includegraphics[width=0.5\linewidth]{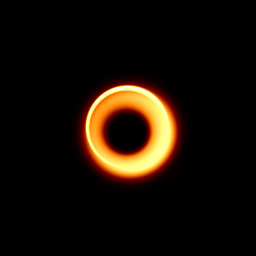}
\caption{Representative ray-traced images of a Kerr black hole surrounded by a geometrically thick, magnetized accretion torus.}
\label{fig0}
\end{figure}
Based on the physical framework described above, ray-tracing techniques can be employed to simulate black-hole images surrounded by geometrically thick accretion tori, leading to the characteristic ``doughnut-like'' intensity morphology \cite{Zhang:2024jrw}, as illustrated in Fig. \ref{fig0}.
This forward modeling approach enables the efficient generation of large ensembles of physically labeled images, which serve as suitable input data for subsequent learning-based inference. 

\bibliography{apssamp}

@ARTICLE{2025ApJS..279...10L,
       author = {{Liu}, Ao and {Zhang}, Zelin and {Chen}, Songbai and {Wen}, Cuihong and {Wang}, Jieci},
        title = "{BCDDM: Branch Correction Denoising Diffusion Model for Black Hole Image Generation}",
      journal = {\apjs},
         year = 2025,
        month = jul,
       volume = {279},
       number = {1},
          eid = {10},
        pages = {10},
          doi = {10.3847/1538-4365/add896},
archivePrefix = {arXiv},
       eprint = {2502.08528},
 primaryClass = {astro-ph.GA},
       adsurl = {https://ui.adsabs.harvard.edu/abs/2025ApJS..279...10L}
}

@ARTICLE{2019ApJ...875L...1E,
       author = {{Event Horizon Telescope Collaboration} and {Akiyama}, Kazunori and {Alberdi}, Antxon and {Alef}, Walter and others},
        title = "{First M87 Event Horizon Telescope Results. I. The Shadow of the Supermassive Black Hole}",
      journal = {\apjl},
         year = 2019,
        month = apr,
       volume = {875},
       number = {1},
          eid = {L1},
        pages = {L1},
          doi = {10.3847/2041-8213/ab0ec7},
archivePrefix = {arXiv},
       eprint = {1906.11238},
 primaryClass = {astro-ph.GA},
       adsurl = {https://ui.adsabs.harvard.edu/abs/2019ApJ...875L...1E}
}

@ARTICLE{2022ApJ...930L..12E,
       author = {{Event Horizon Telescope Collaboration} and {Akiyama}, Kazunori and {Alberdi}, Antxon and {Alef}, Walter and others},
        title = "{First Sagittarius A* Event Horizon Telescope Results. I. The Shadow of the Supermassive Black Hole in the Center of the Milky Way}",
      journal = {\apjl},
         year = 2022,
        month = may,
       volume = {930},
       number = {2},
          eid = {L12},
        pages = {L12},
          doi = {10.3847/2041-8213/ac6674},
       adsurl = {https://ui.adsabs.harvard.edu/abs/2022ApJ...930L..12E}
}

@ARTICLE{2024A&A...681A..79E,
       author = {{Event Horizon Telescope Collaboration} and {Akiyama}, Kazunori and {Alberdi}, Antxon and {Alef}, Walter and others},
        title = "{The persistent shadow of the supermassive black hole of M 87. I. Observations, calibration, imaging, and analysis}",
      journal = {\aap},
         year = 2024,
        month = jan,
       volume = {681},
          eid = {A79},
        pages = {A79},
          doi = {10.1051/0004-6361/202347932},
       adsurl = {https://ui.adsabs.harvard.edu/abs/2024A&A...681A..79E}
}

@ARTICLE{2021ApJ...912...35N,
       author = {{Narayan}, Ramesh and {Palumbo}, Daniel C.~M. and {Johnson}, Michael D. and {Gelles}, Zachary and others},
        title = "{The Polarized Image of a Synchrotron-emitting Ring of Gas Orbiting a Black Hole}",
      journal = {\apj},
         year = 2021,
        month = may,
       volume = {912},
       number = {1},
          eid = {35},
        pages = {35},
          doi = {10.3847/1538-4357/abf117},
archivePrefix = {arXiv},
       eprint = {2105.01804},
 primaryClass = {astro-ph.HE},
       adsurl = {https://ui.adsabs.harvard.edu/abs/2021ApJ...912...35N}
}

@ARTICLE{2018NatAs...2..585M,
       author = {{Mizuno}, Yosuke and {Younsi}, Ziri and {Fromm}, Christian M. and {Porth}, Oliver and others},
        title = "{The current ability to test theories of gravity with black hole shadows}",
      journal = {Nature Astronomy},
         year = 2018,
        month = apr,
       volume = {2},
        pages = {585-590},
          doi = {10.1038/s41550-018-0449-5},
archivePrefix = {arXiv},
       eprint = {1804.05812},
 primaryClass = {astro-ph.GA},
       adsurl = {https://ui.adsabs.harvard.edu/abs/2018NatAs...2..585M}
}

@ARTICLE{2014ARA&A..52..529Y,
       author = {{Yuan}, Feng and {Narayan}, Ramesh},
        title = "{Hot Accretion Flows Around Black Holes}",
      journal = {\araa},
         year = 2014,
        month = aug,
       volume = {52},
        pages = {529-588},
          doi = {10.1146/annurev-astro-082812-141003},
archivePrefix = {arXiv},
       eprint = {1401.0586},
 primaryClass = {astro-ph.HE},
       adsurl = {https://ui.adsabs.harvard.edu/abs/2014ARA&A..52..529Y}
}

@ARTICLE{2022MNRAS.511.3795N,
       author = {{Narayan}, Ramesh and {Chael}, Andrew and {Chatterjee}, Koushik and {Ricarte}, Angelo and others},
        title = "{Jets in magnetically arrested hot accretion flows: geometry, power, and black hole spin-down}",
      journal = {\mnras},
         year = 2022,
        month = apr,
       volume = {511},
       number = {3},
        pages = {3795-3813},
          doi = {10.1093/mnras/stac285},
archivePrefix = {arXiv},
       eprint = {2108.12380},
 primaryClass = {astro-ph.HE},
       adsurl = {https://ui.adsabs.harvard.edu/abs/2022MNRAS.511.3795N}
}

@INPROCEEDINGS{1973blho.conf..215B,
       author = {{Bardeen}, J.~M.},
        title = "{Timelike and null geodesics in the Kerr metric.}",
    booktitle = {Black Holes (Les Astres Occlus)},
         year = 1973,
       editor = {{Dewitt}, C. and {Dewitt}, B.~S.},
        month = jan,
        pages = {215-239},
       adsurl = {https://ui.adsabs.harvard.edu/abs/1973blho.conf..215B}
}

@ARTICLE{2000ApJ...528L..13F,
       author = {{Falcke}, Heino and {Melia}, Fulvio and {Agol}, Eric},
        title = "{Viewing the Shadow of the Black Hole at the Galactic Center}",
      journal = {\apjl},
         year = 2000,
        month = jan,
       volume = {528},
       number = {1},
        pages = {L13-L16},
          doi = {10.1086/312423},
archivePrefix = {arXiv},
       eprint = {astro-ph/9912263},
 primaryClass = {astro-ph},
       adsurl = {https://ui.adsabs.harvard.edu/abs/2000ApJ...528L..13F}
}

@ARTICLE{2020ApJ...897..139B,
       author = {{Broderick}, Avery E. and {Gold}, Roman and {Karami}, Mansour and {Preciado-L{\'o}pez}, Jorge A. and others},
        title = "{THEMIS: A Parameter Estimation Framework for the Event Horizon Telescope}",
      journal = {\apj},
         year = 2020,
        month = jul,
       volume = {897},
       number = {2},
          eid = {139},
        pages = {139},
          doi = {10.3847/1538-4357/ab91a4},
       adsurl = {https://ui.adsabs.harvard.edu/abs/2020ApJ...897..139B}
}

@ARTICLE{2024MNRAS.535.3181Y,
       author = {{Yfantis}, A.~I. and {Zhao}, S. and {Gold}, R. and {Mo{\'s}cibrodzka}, M. and others},
        title = "{Testing Bayesian inference of GRMHD model parameters from VLBI data}",
      journal = {\mnras},
         year = 2024,
        month = dec,
       volume = {535},
       number = {4},
        pages = {3181-3197},
          doi = {10.1093/mnras/stae2509},
archivePrefix = {arXiv},
       eprint = {2409.15417},
 primaryClass = {astro-ph.HE},
       adsurl = {https://ui.adsabs.harvard.edu/abs/2024MNRAS.535.3181Y}
}

@INPROCEEDINGS{2024IAUGA..32P1132C,
       author = {{Chang}, Dominic and {Tiede}, Paul and {Johnson}, Michael and {Palumbo}, Daniel},
        title = "{Bayesian Black Hole Photogrammetry}",
    booktitle = {IAU General Assembly},
         year = 2024,
        month = aug,
          eid = {1132},
        pages = {1132},
          doi = {10.48550/arXiv.2405.04749},
archivePrefix = {arXiv},
       eprint = {2405.04749},
 primaryClass = {astro-ph.HE},
       adsurl = {https://ui.adsabs.harvard.edu/abs/2024IAUGA..32P1132C}
}

@ARTICLE{2026PatRe.17212577Z,
       author = {{Zhang}, Zhizheng and {Shao}, Zhenfeng and {Ma}, Jiayi and {Zhang}, Jindou and {Wang}, Yu and {Liao}, Zhenghao and {Cheng}, Gui and {Liu}, Jun and {Guo}, Mingqiang and {Wu}, Liang},
        title = "{Land-cover prior diffusion probabilistic model for remote sensing image super resolution}",
      journal = {Pattern Recognition},
         year = 2026,
        month = apr,
       volume = {172},
          eid = {112577},
        pages = {112577},
          doi = {10.1016/j.patcog.2025.112577},
       adsurl = {https://ui.adsabs.harvard.edu/abs/2026PatRe.17212577Z}
}

@ARTICLE{2026PatRe.17112232L,
       author = {{Liu}, Shuaixin and {Li}, Kunqian and {Ding}, Yilin and {Qi}, Qi},
        title = "{Underwater image enhancement by diffusion model with customized CLIP-classifier}",
      journal = {Pattern Recognition},
         year = 2026,
        month = mar,
       volume = {171},
          eid = {112232},
        pages = {112232},
          doi = {10.1016/j.patcog.2025.112232},
       adsurl = {https://ui.adsabs.harvard.edu/abs/2026PatRe.17112232L}
}

@ARTICLE{2026PatRe.17112081X,
       author = {{Xie}, Dirui and {Hu}, Xiaofang and {Zhou}, Yue and {Duan}, Shukai},
        title = "{All-in-one adverse weather removal via dual state space-based diffusion model with degradation-aware guidance}",
      journal = {Pattern Recognition},
         year = 2026,
        month = mar,
       volume = {171},
          eid = {112081},
        pages = {112081},
          doi = {10.1016/j.patcog.2025.112081},
       adsurl = {https://ui.adsabs.harvard.edu/abs/2026PatRe.17112081X}
}

@ARTICLE{2026Meas..25919599O,
       author = {{Oh}, Joowon and {Lee}, Jeaho},
        title = "{Mitigating ANC pressure effect in Active Noise Cancellation using diffusion-based generative models}",
      journal = {Measurement},
         year = 2026,
        month = feb,
       volume = {259},
          eid = {119599},
        pages = {119599},
          doi = {10.1016/j.measurement.2025.119599},
       adsurl = {https://ui.adsabs.harvard.edu/abs/2026Meas..25919599O}
}

@ARTICLE{2026PatRe.16911934C,
       author = {{Chang}, Ziyi and {Koulieris}, George A. and {Chang}, Hyung Jin and {Shum}, Hubert P.~H.},
        title = "{On the design fundamentals of diffusion models: A survey}",
      journal = {Pattern Recognition},
         year = 2026,
        month = jan,
       volume = {169},
          eid = {111934},
        pages = {111934},
          doi = {10.1016/j.patcog.2025.111934},
archivePrefix = {arXiv},
       eprint = {2306.04542},
 primaryClass = {stat.ML},
       adsurl = {https://ui.adsabs.harvard.edu/abs/2026PatRe.16911934C}
}

@ARTICLE{2018ApJ...864....7M,
       author = {{Medeiros}, Lia and {Lauer}, Tod R. and {Psaltis}, Dimitrios and {{\"O}zel}, Feryal},
        title = "{Principal Component Analysis as a Tool for Characterizing Black Hole Images and Variability}",
      journal = {\apj},
         year = 2018,
        month = sep,
       volume = {864},
       number = {1},
          eid = {7},
        pages = {7},
          doi = {10.3847/1538-4357/aad37a},
archivePrefix = {arXiv},
       eprint = {1804.05903},
 primaryClass = {astro-ph.IM},
       adsurl = {https://ui.adsabs.harvard.edu/abs/2018ApJ...864....7M}
}

@ARTICLE{2022ApJ...927..111H,
       author = {{Hallur}, Pravita and {Medeiros}, Lia and {Lauer}, Tod R.},
        title = "{A Red-noise Eigenbasis for the Reconstruction of Blobby Images}",
      journal = {\apj},
         year = 2022,
        month = mar,
       volume = {927},
       number = {1},
          eid = {111},
        pages = {111},
          doi = {10.3847/1538-4357/ac502a},
archivePrefix = {arXiv},
       eprint = {2111.01168},
 primaryClass = {astro-ph.IM},
       adsurl = {https://ui.adsabs.harvard.edu/abs/2022ApJ...927..111H}
}

@ARTICLE{2025ApJ...984...86P,
       author = {{Psaltis}, Dimitrios and {{\"O}zel}, Feryal and {Medeiros}, Lia and {Lauer}, Tod R.},
        title = "{Theoretical Foundation of Black Hole Image Reconstruction Using PRIMO}",
      journal = {\apj},
         year = 2025,
        month = may,
       volume = {984},
       number = {1},
          eid = {86},
        pages = {86},
          doi = {10.3847/1538-4357/ada60f},
archivePrefix = {arXiv},
       eprint = {2408.10322},
 primaryClass = {astro-ph.IM},
       adsurl = {https://ui.adsabs.harvard.edu/abs/2025ApJ...984...86P}
}

@ARTICLE{2013arXiv1310.0425F,
       author = {{Fefferman}, Charles and {Mitter}, Sanjoy and {Narayanan}, Hariharan},
        title = "{Testing the Manifold Hypothesis}",
      journal = {arXiv e-prints},
         year = 2013,
        month = oct,
          eid = {arXiv:1310.0425},
        pages = {arXiv:1310.0425},
          doi = {10.48550/arXiv.1310.0425},
archivePrefix = {arXiv},
       eprint = {1310.0425},
 primaryClass = {math.ST},
       adsurl = {https://ui.adsabs.harvard.edu/abs/2013arXiv1310.0425F}
}

@ARTICLE{6789755,
  author={Belkin, Mikhail and Niyogi, Partha},
  journal={Neural Computation}, 
  title={Laplacian Eigenmaps for Dimensionality Reduction and Data Representation}, 
  year={2003},
  volume={15},
  number={6},
  pages={1373-1396},
  doi={10.1162/089976603321780317}
}

@ARTICLE{2012arXiv1206.5538B,
       author = {{Bengio}, Yoshua and {Courville}, Aaron and {Vincent}, Pascal},
        title = "{Representation Learning: A Review and New Perspectives}",
      journal = {arXiv e-prints},
         year = 2012,
        month = jun,
          eid = {arXiv:1206.5538},
        pages = {arXiv:1206.5538},
          doi = {10.48550/arXiv.1206.5538},
archivePrefix = {arXiv},
       eprint = {1206.5538},
 primaryClass = {cs.LG},
       adsurl = {https://ui.adsabs.harvard.edu/abs/2012arXiv1206.5538B}
}

@ARTICLE{2021arXiv210408894P,
       author = {{Pope}, Phillip and {Zhu}, Chen and {Abdelkader}, Ahmed and {Goldblum}, Micah and {Goldstein}, Tom},
        title = "{The Intrinsic Dimension of Images and Its Impact on Learning}",
      journal = {arXiv e-prints},
         year = 2021,
        month = apr,
          eid = {arXiv:2104.08894},
        pages = {arXiv:2104.08894},
          doi = {10.48550/arXiv.2104.08894},
archivePrefix = {arXiv},
       eprint = {2104.08894},
 primaryClass = {cs.CV},
       adsurl = {https://ui.adsabs.harvard.edu/abs/2021arXiv210408894P}
}

@article{Komissarov:2006nz,
    author = "Komissarov, S. S.",
    title = "{Magnetized Tori around Kerr Black Holes: Analytic Solutions with a Toroidal Magnetic Field}",
    eprint = "astro-ph/0601678",
    archivePrefix = "arXiv",
    doi = "10.1111/j.1365-2966.2006.10183.x",
    journal = "Mon. Not. Roy. Astron. Soc.",
    volume = "368",
    pages = "993--1000",
    year = "2006"
}

@article{Yuan:2014gma,
    author = "Yuan, Feng and Narayan, Ramesh",
    title = "{Hot Accretion Flows Around Black Holes}",
    eprint = "1401.0586",
    archivePrefix = "arXiv",
    primaryClass = "astro-ph.HE",
    doi = "10.1146/annurev-astro-082812-141003",
    journal = "Ann. Rev. Astron. Astrophys.",
    volume = "52",
    pages = "529--588",
    year = "2014"
}

@ARTICLE{Zhang:2024lsf,
       author = {{Zhang}, Zhenyu and {Hou}, Yehui and {Guo}, Minyong and {Chen}, Bin},
        title = "{Imaging thick accretion disks and jets surrounding black holes}",
      journal = {Journal of Cosmology and Astroparticle Physics},
         year = 2024,
        month = may,
       volume = {2024},
       number = {5},
          eid = {032},
        pages = {032},
          doi = {10.1088/1475-7516/2024/05/032},
archivePrefix = {arXiv},
       eprint = {2401.14794},
 primaryClass = {astro-ph.HE},
       adsurl = {https://ui.adsabs.harvard.edu/abs/2024JCAP...05..032Z}
}

@ARTICLE{Hou:2023bep,
       author = {{Hou}, Yehui and {Zhang}, Zhenyu and {Guo}, Minyong and {Chen}, Bin},
        title = "{A new analytical model of magnetofluids surrounding rotating black holes}",
      journal = {Journal of Cosmology and Astroparticle Physics},
         year = 2024,
        month = feb,
       volume = {2024},
       number = {2},
          eid = {030},
        pages = {030},
          doi = {10.1088/1475-7516/2024/02/030},
archivePrefix = {arXiv},
       eprint = {2309.13304},
 primaryClass = {gr-qc},
       adsurl = {https://ui.adsabs.harvard.edu/abs/2024JCAP...02..030H}
}

@ARTICLE{Zhang:2024jrw,
       author = {{Zhang}, Zelin and {Chen}, Songbai and {Jing}, Jiliang},
        title = "{Images of Kerr-MOG black holes surrounded by geometrically thick magnetized equilibrium tori}",
      journal = {Journal of Cosmology and Astroparticle Physics},
         year = 2024,
        month = sep,
       volume = {2024},
       number = {9},
          eid = {027},
        pages = {027},
          doi = {10.1088/1475-7516/2024/09/027},
archivePrefix = {arXiv},
       eprint = {2404.12223},
 primaryClass = {gr-qc},
       adsurl = {https://ui.adsabs.harvard.edu/abs/2024JCAP...09..027Z}
}

@dataset{Zhang111,
  author       = {Zhang, Zelin and
                  Chen, Songbai},
  title        = {Supplementary data for "BCDDM: Branch-Corrected
                   Denoising Diffusion Model for Black Hole Image
                   Generation"
                  },
  month        = apr,
  year         = 2025,
  publisher    = {Zenodo},
  doi          = {10.5281/zenodo.15354648},
  url          = {https://doi.org/10.5281/zenodo.15354648},
}

@ARTICLE{Liu:2025wwq,
       author = {{Liu}, Wentao and {Liu}, Yang and {Wu}, Di and {Liu}, Yu-Xiao},
        title = "{A Universal Framework for Horizon-Scale Tests of Gravity with Black Hole Shadows}",
      journal = {arXiv e-prints},
         year = 2025,
        month = nov,
          eid = {arXiv:2511.06017},
        pages = {arXiv:2511.06017},
          doi = {10.48550/arXiv.2511.06017},
archivePrefix = {arXiv},
       eprint = {2511.06017},
 primaryClass = {gr-qc},
       adsurl = {https://ui.adsabs.harvard.edu/abs/2025arXiv251106017L}
}

@article{Cunha:2018acu,
    author = "Cunha, Pedro V. P. and Herdeiro, Carlos A. R.",
    title = "{Shadows and strong gravitational lensing: a brief review}",
    eprint = "1801.00860",
    archivePrefix = "arXiv",
    primaryClass = "gr-qc",
    doi = "10.1007/s10714-018-2361-9",
    journal = "Gen. Rel. Grav.",
    volume = "50",
    number = "4",
    pages = "42",
    year = "2018"
}

@article{Gralla:2019xty,
    author = "Gralla, Samuel E. and Holz, Daniel E. and Wald, Robert M.",
    title = "{Black Hole Shadows, Photon Rings, and Lensing Rings}",
    eprint = "1906.00873",
    archivePrefix = "arXiv",
    primaryClass = "astro-ph.HE",
    doi = "10.1103/PhysRevD.100.024018",
    journal = "Phys. Rev. D",
    volume = "100",
    number = "2",
    pages = "024018",
    year = "2019"
}

@article{Cunha:2015yba,
    author = "Cunha, Pedro V. P. and Herdeiro, Carlos A. R. and Radu, Eugen and Runarsson, Helgi F.",
    title = "{Shadows of Kerr black holes with scalar hair}",
    eprint = "1509.00021",
    archivePrefix = "arXiv",
    primaryClass = "gr-qc",
    doi = "10.1103/PhysRevLett.115.211102",
    journal = "Phys. Rev. Lett.",
    volume = "115",
    number = "21",
    pages = "211102",
    year = "2015"
}

@article{Cunha:2019dwb,
    author = "Cunha, Pedro V. P. and Herdeiro, Carlos A. R. and Radu, Eugen",
    title = "{Spontaneously Scalarized Kerr Black Holes in Extended Scalar-Tensor{\textendash}Gauss-Bonnet Gravity}",
    eprint = "1904.09997",
    archivePrefix = "arXiv",
    primaryClass = "gr-qc",
    doi = "10.1103/PhysRevLett.123.011101",
    journal = "Phys. Rev. Lett.",
    volume = "123",
    number = "1",
    pages = "011101",
    year = "2019"
}

@article{Hu:2020usx,
    author = "Hu, Zezhou and Zhong, Zhen and Li, Peng-Cheng and Guo, Minyong and Chen, Bin",
    title = "{QED effect on a black hole shadow}",
    eprint = "2012.07022",
    archivePrefix = "arXiv",
    primaryClass = "gr-qc",
    doi = "10.1103/PhysRevD.103.044057",
    journal = "Phys. Rev. D",
    volume = "103",
    number = "4",
    pages = "044057",
    year = "2021"
}

@article{Bacchini:2021fig,
    author = "Bacchini, Fabio and Mayerson, Daniel R. and Ripperda, Bart and Davelaar, Jordy and Olivares, H{\'e}ctor and Hertog, Thomas and Vercnocke, Bert",
    title = "{Fuzzball Shadows: Emergent Horizons from Microstructure}",
    eprint = "2103.12075",
    archivePrefix = "arXiv",
    primaryClass = "hep-th",
    doi = "10.1103/PhysRevLett.127.171601",
    journal = "Phys. Rev. Lett.",
    volume = "127",
    number = "17",
    pages = "171601",
    year = "2021"
}

@ARTICLE{Zhang:2025xnl,
       author = {{Zhang}, Yu-Peng and {Wei}, Shao-Wen and {Liu}, Yu-Xiao},
        title = "{Emerging black hole shadow from collapsing boson star}",
      journal = {arXiv e-prints},
         year = 2025,
        month = mar,
          eid = {arXiv:2503.14159},
        pages = {arXiv:2503.14159},
          doi = {10.48550/arXiv.2503.14159},
archivePrefix = {arXiv},
       eprint = {2503.14159},
 primaryClass = {gr-qc},
       adsurl = {https://ui.adsabs.harvard.edu/abs/2025arXiv250314159Z}
}

@article{Perlick:2021aok,
    author = "Perlick, Volker and Tsupko, Oleg Yu.",
    title = "{Calculating black hole shadows: Review of analytical studies}",
    eprint = "2105.07101",
    archivePrefix = "arXiv",
    primaryClass = "gr-qc",
    doi = "10.1016/j.physrep.2021.10.004",
    journal = "Phys. Rept.",
    volume = "947",
    pages = "1--39",
    year = "2022"
}

@article{Bronzwaer:2021lzo,
    author = "Bronzwaer, Thomas and Falcke, Heino",
    title = "{The Nature of Black Hole Shadows}",
    eprint = "2108.03966",
    archivePrefix = "arXiv",
    primaryClass = "astro-ph.HE",
    doi = "10.3847/1538-4357/ac1738",
    journal = "Astrophys. J.",
    volume = "920",
    number = "2",
    pages = "155",
    year = "2021"
}

@article{Konoplya:2021slg,
    author = "Konoplya, R. A. and Zhidenko, A.",
    title = "{Shadows of parametrized axially symmetric black holes allowing for separation of variables}",
    eprint = "2103.03855",
    archivePrefix = "arXiv",
    primaryClass = "gr-qc",
    doi = "10.1103/PhysRevD.103.104033",
    journal = "Phys. Rev. D",
    volume = "103",
    number = "10",
    pages = "104033",
    year = "2021"
}

@article{Vagnozzi:2022moj,
    author = "Vagnozzi, Sunny and others",
    title = "{Horizon-scale tests of gravity theories and fundamental physics from the Event Horizon Telescope image of Sagittarius A}",
    eprint = "2205.07787",
    archivePrefix = "arXiv",
    primaryClass = "gr-qc",
    reportNumber = "UCI-HEP-TR-2022-07",
    doi = "10.1088/1361-6382/acd97b",
    journal = "Class. Quant. Grav.",
    volume = "40",
    number = "16",
    pages = "165007",
    year = "2023"
}

@article{Bambi:2019tjh,
    author = "Bambi, Cosimo and Freese, Katherine and Vagnozzi, Sunny and Visinelli, Luca",
    title = "{Testing the rotational nature of the supermassive object M87* from the circularity and size of its first image}",
    eprint = "1904.12983",
    archivePrefix = "arXiv",
    primaryClass = "gr-qc",
    doi = "10.1103/PhysRevD.100.044057",
    journal = "Phys. Rev. D",
    volume = "100",
    number = "4",
    pages = "044057",
    year = "2019"
}

@article{Chen:2022scf,
    author = "Chen, Songbai and Jing, Jiliang and Qian, Wei-Liang and Wang, Bin",
    title = "{Black hole images: A review}",
    eprint = "2301.00113",
    archivePrefix = "arXiv",
    primaryClass = "astro-ph.HE",
    doi = "10.1007/s11433-022-2059-5",
    journal = "Sci. China Phys. Mech. Astron.",
    volume = "66",
    number = "6",
    pages = "260401",
    year = "2023"
}

@article{Ayzenberg:2023hfw,
    author = "Ayzenberg, D. and others",
    title = "{Fundamental physics opportunities with future ground-based mm/sub-mm VLBI arrays}",
    eprint = "2312.02130",
    archivePrefix = "arXiv",
    primaryClass = "astro-ph.HE",
    doi = "10.1007/s41114-025-00057-0",
    journal = "Living Rev. Rel.",
    volume = "28",
    number = "1",
    pages = "4",
    year = "2025",
    note = "[Erratum: Living Rev.Rel. 28, 7 (2025)]"
}

@article{Crinquand:2022ogr,
    author = "Crinquand, Benjamin and Cerutti, Beno{\^\i}t and Dubus, Guillaume and Parfrey, Kyle and Philippov, Alexander A.",
    title = "{Synthetic Images of Magnetospheric Reconnection-Powered Radiation around Supermassive Black Holes}",
    eprint = "2202.04472",
    archivePrefix = "arXiv",
    primaryClass = "astro-ph.HE",
    doi = "10.1103/PhysRevLett.129.205101",
    journal = "Phys. Rev. Lett.",
    volume = "129",
    number = "20",
    pages = "205101",
    year = "2022"
}

@article{Chen:2024nua,
    author = "Chen, Yifan and Ding, Ran and Liu, Yuxin and Mizuno, Yosuke and Shu, Jing and Yu, Haiyue and Zeng, Yanjie",
    title = "{Illuminating Black Hole Shadows with Dark Matter Annihilation}",
    eprint = "2404.16673",
    archivePrefix = "arXiv",
    primaryClass = "hep-ph",
    doi = "10.1103/yxqg-363n",
    journal = "Phys. Rev. Lett.",
    volume = "135",
    number = "12",
    pages = "121001",
    year = "2025"
}

@article{Uniyal:2025uvc,
    author = "Uniyal, Akhil and Dihingia, Indu K. and Mizuno, Yosuke and Rezzolla, Luciano",
    title = "{The future ability to test theories of gravity with black-hole shadows}",
    eprint = "2511.03789",
    archivePrefix = "arXiv",
    primaryClass = "gr-qc",
    doi = "10.1038/s41550-025-02695-4",
    journal = "Nature Astron.",
    volume = "1",
    pages = "8",
    year = "2025"
}

@article{Liu:2025lwj,
    author = "Liu, Wentao and Huang, Hongxia and Wu, Di and Wang, Jieci",
    title = "{Lorentz violation signatures in the low-energy sector of Ho{\v{r}}ava gravity from black hole shadow observations}",
    eprint = "2506.13504",
    archivePrefix = "arXiv",
    primaryClass = "gr-qc",
    doi = "10.1016/j.physletb.2025.139812",
    journal = "Phys. Lett. B",
    volume = "868",
    pages = "139812",
    year = "2025"
}

@article{Liu:2024soc,
    author = "Liu, Wentao and Wu, Di and Wang, Jieci",
    title = "{Light rings and shadows of static black holes in effective quantum gravity}",
    eprint = "2408.05569",
    archivePrefix = "arXiv",
    primaryClass = "gr-qc",
    doi = "10.1016/j.physletb.2024.139052",
    journal = "Phys. Lett. B",
    volume = "858",
    pages = "139052",
    year = "2024"
}

@article{Liu:2024iec,
    author = "Liu, Wentao and Wu, Di and Wang, Jieci",
    title = "{Light rings and shadows of static black holes in effective quantum gravity II: A new solution without Cauchy horizons}",
    eprint = "2412.18083",
    archivePrefix = "arXiv",
    primaryClass = "gr-qc",
    doi = "10.1016/j.physletb.2025.139742",
    journal = "Phys. Lett. B",
    volume = "868",
    pages = "139742",
    year = "2025"
}

@ARTICLE{Liu:2024lbi,
       author = {{Liu}, Wentao and {Wu}, Di and {Fang}, Xiongjun and {Jing}, Jiliang and {Wang}, Jieci},
        title = "{Kerr-MOG-(A)dS black hole and its shadow in scalar-tensor-vector gravity theory}",
      journal = {JCAP},
         year = 2024,
        month = aug,
       volume = {2024},
       number = {8},
          eid = {035},
        pages = {035},
          doi = {10.1088/1475-7516/2024/08/035},
archivePrefix = {arXiv},
       eprint = {2406.00579},
 primaryClass = {gr-qc},
       adsurl = {https://ui.adsabs.harvard.edu/abs/2024JCAP...08..035L}
}

@ARTICLE{Liu:2024lve,
       author = {{Liu}, Wentao and {Wu}, Di and {Wang}, Jieci},
        title = "{Shadow of slowly rotating Kalb-Ramond black holes}",
      journal = {Journal of Cosmology and Astroparticle Physics},
         year = 2025,
        month = may,
       volume = {2025},
       number = {5},
          eid = {017},
        pages = {017},
          doi = {10.1088/1475-7516/2025/05/017},
archivePrefix = {arXiv},
       eprint = {2407.07416},
 primaryClass = {gr-qc},
       adsurl = {https://ui.adsabs.harvard.edu/abs/2025JCAP...05..017L}
}

@ARTICLE{Zhu:2025ouf,
       author = {{Zhu}, Qing-Hua},
        title = "{Ray tracing for the Terrell-Penrose effect in black hole spacetime}",
      journal = {arXiv e-prints},
         year = 2025,
        month = nov,
          eid = {arXiv:2511.09355},
        pages = {arXiv:2511.09355},
          doi = {10.48550/arXiv.2511.09355},
archivePrefix = {arXiv},
       eprint = {2511.09355},
 primaryClass = {gr-qc},
       adsurl = {https://ui.adsabs.harvard.edu/abs/2025arXiv251109355Z}
}

@article{Zhu:2024vxw,
    author = "Zhu, Qing-Hua",
    title = "{Observational signatures from higher-order images of moving hotspots in accretion disks}",
    eprint = "2411.04001",
    archivePrefix = "arXiv",
    primaryClass = "gr-qc",
    doi = "10.1103/PhysRevD.111.044010",
    journal = "Phys. Rev. D",
    volume = "111",
    number = "4",
    pages = "044010",
    year = "2025"
}

@article{Zhu:2025jqh,
    author = "Zhu, Qing-Hua",
    title = "{Auto- and cross-correlations for multiple images of corotating hotspots in accretion disks}",
    eprint = "2503.22343",
    archivePrefix = "arXiv",
    primaryClass = "astro-ph.HE",
    doi = "10.1103/gddn-hzpv",
    journal = "Phys. Rev. D",
    volume = "112",
    number = "6",
    pages = "064021",
    year = "2025"
}

@article{Zhu:2023kei,
    author = "Zhu, Qing-Hua",
    title = "{Aberration effect on lower-order images of a thin accretion disk in the astrometric approach}",
    eprint = "2311.17390",
    archivePrefix = "arXiv",
    primaryClass = "gr-qc",
    doi = "10.1103/PhysRevD.109.044057",
    journal = "Phys. Rev. D",
    volume = "109",
    number = "4",
    pages = "044057",
    year = "2024"
}

@ARTICLE{Qiao:2025tnx,
       author = {{Qiao}, Chenkai and {Li}, Ming and {Xie}, Donghui and {Guo}, Minyong},
        title = "{Geometric Approach to Light Rings in Axially Symmetric Spacetimes}",
      journal = {arXiv e-prints},
         year = 2025,
        month = dec,
          eid = {arXiv:2512.20802},
        pages = {arXiv:2512.20802},
          doi = {10.48550/arXiv.2512.20802},
archivePrefix = {arXiv},
       eprint = {2512.20802},
 primaryClass = {gr-qc},
       adsurl = {https://ui.adsabs.harvard.edu/abs/2025arXiv251220802Q}
}

@ARTICLE{Wan:2025gbm,
       author = {{Wan}, Qian and {Hou}, Yehui and {Guo}, Minyong},
        title = "{Probing the Scalar Hair of Rotating Horndeski Black Holes through Thick Disk Images}",
      journal = {arXiv e-prints},
         year = 2025,
        month = nov,
          eid = {arXiv:2512.00917},
        pages = {arXiv:2512.00917},
          doi = {10.48550/arXiv.2512.00917},
archivePrefix = {arXiv},
       eprint = {2512.00917},
 primaryClass = {gr-qc},
       adsurl = {https://ui.adsabs.harvard.edu/abs/2025arXiv251200917W}
}

@article{Hou:2024qqo,
    author = "Hou, Yehui and Huang, Jiewei and Guo, Minyong and Mizuno, Yosuke and Chen, Bin",
    title = "{Near-horizon Polarization as a Diagnostic of Black Hole Spacetime}",
    eprint = "2409.07248",
    archivePrefix = "arXiv",
    primaryClass = "gr-qc",
    reportNumber = "988 L51",
    doi = "10.3847/2041-8213/adee09",
    journal = "Astrophys. J. Lett.",
    volume = "988",
    number = "2",
    pages = "L51",
    year = "2025"
}

@ARTICLE{Cupp:2025zwv,
       author = {{Cupp}, Samuel and {Werneck}, Leonardo R. and {Jacques}, Terrence Pierre and {Tootle}, Samuel and {Etienne}, Zachariah B.},
        title = "{GRHayL: a modern, infrastructure-agnostic, extensible library for GRMHD simulations}",
      journal = {arXiv e-prints},
         year = 2025,
        month = dec,
          eid = {arXiv:2512.15846},
        pages = {arXiv:2512.15846},
          doi = {10.48550/arXiv.2512.15846},
archivePrefix = {arXiv},
       eprint = {2512.15846},
 primaryClass = {gr-qc},
       adsurl = {https://ui.adsabs.harvard.edu/abs/2025arXiv251215846C}
}

@article{Abramowicz:2011xu,
    author = "Abramowicz, Marek A. and Fragile, P. Chris",
    title = "{Foundations of Black Hole Accretion Disk Theory}",
    eprint = "1104.5499",
    archivePrefix = "arXiv",
    primaryClass = "astro-ph.HE",
    reportNumber = "NSF-KITP-12-055",
    doi = "10.12942/lrr-2013-1",
    journal = "Living Rev. Rel.",
    volume = "16",
    pages = "1",
    year = "2013"
}

@article{EventHorizonTelescope:2019pcy,
    author = "Porth, Oliver and others",
    collaboration = "Event Horizon Telescope",
    title = "{The Event Horizon General Relativistic Magnetohydrodynamic Code Comparison Project}",
    eprint = "1904.04923",
    archivePrefix = "arXiv",
    primaryClass = "astro-ph.HE",
    doi = "10.3847/1538-4365/ab29fd",
    journal = "Astrophys. J. Suppl.",
    volume = "243",
    number = "2",
    pages = "26",
    year = "2019"
}

@ARTICLE{Aslam:2025hgl,
       author = {{Israr Aslam}, Muhammad and {Saleem}, Rabia and {Yang}, Chen-Yu and {Zeng}, Xiao-Xiong},
        title = "{Imprints of Dark Matter on the Shadow and Polarization Images of a Black Hole Illuminated by Various Thick Disks}",
      journal = {arXiv e-prints},
         year = 2025,
        month = nov,
          eid = {arXiv:2512.00964},
        pages = {arXiv:2512.00964},
          doi = {10.48550/arXiv.2512.00964},
archivePrefix = {arXiv},
       eprint = {2512.00964},
 primaryClass = {gr-qc},
       adsurl = {https://ui.adsabs.harvard.edu/abs/2025arXiv251200964I}
}

@ARTICLE{Wang:2025qpv,
       author = {{Wang}, Xinyu and {Ye}, Huan and {Zeng}, Xiao-Xiong},
        title = "{Imaging and polarization patterns of various thick disks around Kerr-MOG black holes}",
      journal = {arXiv e-prints},
         year = 2025,
        month = nov,
          eid = {arXiv:2511.09379},
        pages = {arXiv:2511.09379},
          doi = {10.48550/arXiv.2511.09379},
archivePrefix = {arXiv},
       eprint = {2511.09379},
 primaryClass = {gr-qc},
       adsurl = {https://ui.adsabs.harvard.edu/abs/2025arXiv251109379W}
}

@ARTICLE{Zeng:2025pch,
       author = {{Zeng}, Xiao-Xiong and {Yang}, Chen-Yu and {Israr Aslam}, M. and {Saleem}, Rabia},
        title = "{Probing Horndeski Gravity via Kerr Black Hole: Insights from Thin Accretion Disks and Shadows with EHT Observations}",
      journal = {arXiv e-prints},
         year = 2025,
        month = sep,
          eid = {arXiv:2509.05803},
        pages = {arXiv:2509.05803},
          doi = {10.48550/arXiv.2509.05803},
archivePrefix = {arXiv},
       eprint = {2509.05803},
 primaryClass = {gr-qc},
       adsurl = {https://ui.adsabs.harvard.edu/abs/2025arXiv250905803Z}
}

@article{Zeng:2021dlj,
    author = "Zeng, Xiao-Xiong and Li, Guo-Ping and He, Ke-Jian",
    title = "{The shadows and observational appearance of a noncommutative black hole surrounded by various profiles of accretions}",
    eprint = "2106.14478",
    archivePrefix = "arXiv",
    primaryClass = "hep-th",
    doi = "10.1016/j.nuclphysb.2021.115639",
    journal = "Nucl. Phys. B",
    volume = "974",
    pages = "115639",
    year = "2022"
}

@article{Wang:2025dfn,
    author = "Wang, Mingzhi and Zhang, Cheng-Yong and Chen, Songbai and Jing, Jiliang",
    title = "{Dynamic shadow of a black hole with a self-interacting massive complex scalar hair*}",
    eprint = "2507.20569",
    archivePrefix = "arXiv",
    primaryClass = "gr-qc",
    doi = "10.1088/1674-1137/ae1442",
    journal = "Chin. Phys. C",
    volume = "50",
    pages = "025102",
    year = "2025"
}

@ARTICLE{Huang:2025xqd,
       author = {{Huang}, Yang and {Liu}, Dao-Jun and {Zhang}, Hongsheng},
        title = "{Dynamics of photons and shadows for black holes haired with parity-odd fields}",
      journal = {Journal of High Energy Physics},
         year = 2025,
        month = nov,
       volume = {2025},
       number = {11},
          eid = {27},
        pages = {27},
          doi = {10.1007/JHEP11(2025)027},
archivePrefix = {arXiv},
       eprint = {2507.04761},
 primaryClass = {gr-qc},
       adsurl = {https://ui.adsabs.harvard.edu/abs/2025JHEP...11..027H}
}

@article{Huang:2024gtu,
    author = "Huang, Yang and Liu, Dao-Jun and Zhang, Hongsheng",
    title = "{Lensing and light rings of parity-odd rotating boson stars}",
    eprint = "2410.20867",
    archivePrefix = "arXiv",
    primaryClass = "gr-qc",
    doi = "10.1007/s11433-024-2577-0",
    journal = "Sci. China Phys. Mech. Astron.",
    volume = "68",
    number = "8",
    pages = "280411",
    year = "2025"
}

@article{Gao:2023mjb,
    author = "Gao, Xiao-Jun and Sui, Tao-Tao and Zeng, Xiao-Xiong and An, Yu-Sen and Hu, Ya-Peng",
    title = "{Investigating shadow images and rings of the charged Horndeski black hole illuminated by various thin accretions}",
    eprint = "2311.11780",
    archivePrefix = "arXiv",
    primaryClass = "gr-qc",
    doi = "10.1140/epjc/s10052-023-12231-1",
    journal = "Eur. Phys. J. C",
    volume = "83",
    pages = "1052",
    year = "2023"
}

@article{Sui:2023tje,
    author = "Sui, Tao-Tao and Wang, Zi-Liang and Guo, Wen-Di",
    title = "{The effect of scalar hair on the charged black hole with the images from accretions disk}",
    eprint = "2311.10946",
    archivePrefix = "arXiv",
    primaryClass = "gr-qc",
    doi = "10.1140/epjc/s10052-024-12807-5",
    journal = "Eur. Phys. J. C",
    volume = "84",
    number = "4",
    pages = "441",
    year = "2024"
}

@ARTICLE{2025PhRvD.111f3523P,
       author = {{Park}, Minsu and {Gatti}, Marco and {Jain}, Bhuvnesh},
        title = "{Dimensionality reduction techniques for statistical inference in cosmology}",
      journal = {\prd},
         year = 2025,
        month = mar,
       volume = {111},
       number = {6},
          eid = {063523},
        pages = {063523},
          doi = {10.1103/PhysRevD.111.063523},
archivePrefix = {arXiv},
       eprint = {2409.02102},
 primaryClass = {astro-ph.CO},
       adsurl = {https://ui.adsabs.harvard.edu/abs/2025PhRvD.111f3523P}
}

@book{bishop2006pattern,
  title={Pattern Recognition and Machine Learning},
  author={Bishop, Christopher M.},
  year={2006},
  publisher={Springer}
}

@book{goodfellow2016deep,
  title={Deep Learning},
  author={Goodfellow, Ian and Bengio, Yoshua and Courville, Aaron},
  year={2016},
  publisher={MIT Press}
}

@ARTICLE{vaswani2017attention,
       author = {Vaswani, Ashish and Shazeer, Noam and Parmar, Niki and Uszkoreit, Jakob and Jones, Llion and Gomez, Aidan N and Kaiser, {\L}ukasz and Polosukhin, Illia},
        title = "{Attention is all you need}",
      journal = {Advances in neural information processing systems},
         year = 2017,
       volume = {30},
archivePrefix = {arXiv},
       eprint = {1706.03762},
 primaryClass = {cs.CL},
       adsurl = {https://arxiv.org/abs/1706.03762},
}

\end{document}